\documentclass[twocolumn]{svjour3}         % twocolumn
\smartqed  % flush right qed marks, e.g. at end of proof
\usepackage{graphicx}
\begin{document}

\title{Detecting the Most Distant (z$>$7) Objects with ALMA \thanks{
    It is our pleasure to thank our collaborators on this project:
    Frank Bertoldi, Dominik Riechers, Pierre Cox, Roberto Maiolino and
    Axel Wei\ss.}  } \subtitle{}


\author{Fabian Walter and Chris Carilli}


\institute{F.~Walter \at
              Max Planck Insitut f\"ur Astronomie Heidelberg,
              \email{walter@mpia.de}             \\
           C.~Carilli \at National Radio Astronomy Observatory, 
              \email{ccarilli@nrao.edu}
}

\date{Received: date / Accepted: date}

\maketitle

\begin{abstract}

  Detecting and studying objects at the highest redshifts, out to the
  end of Cosmic Reionization at z$>$7, is clearly a key science goal
  of ALMA. ALMA will in principle be able to detect objects in this
  redshift range both from high-J (J$>$7) CO transitions and emission
  from ionized carbon, [CII], which is one of the main cooling lines
  of the ISM.  ALMA will even be able to resolve this emission for
  individual targets, which will be one of the few ways to determine
  dynamical masses for systems in the Epoch of Reionization.  We
  discuss some of the current problems regarding the detection and
  characterization of objects at high redshifts and how ALMA will
  eliminate most (but not all) of them.

% \keywords{First keyword \and Second keyword \and More}
\end{abstract}

\section{Introduction: The highest redshift galaxies}
\vspace*{-2.5mm}

In recent years, deep narrow band surveys have revealed a major
population of Lyman Alpha Emitters (LAE) out to very high redshifts
(e.g.  Hu et al.\ 2002, Kurk et al.\ 2004, Stern et al.\ 2005,
Murayama et al.\ 2007). In particular, Taniguchi et al.\ (2005) report
the detection of 9 spectroscopically confirmed LAE at redshifts of
z$\sim$6.6 in the Subaru Deep Field (currently, the published LAE
redshift record holder is at z=6.98, Iye et al.\ 2006).  The mere
presence of Lyman alpha emission in these sources provides strong
evidence that they are undergoing bursts of star formation: the star
formation rates of individual objects are
$\sim$10\,M$_{\odot}$\,yr$^{-1}$ (based on their FUV luminosities) and
their redshifts place them well within the end of cosmic reionization.
They also appear to be very numerous: Tanaguchi et al.\ 2005 find
$\sim$30 LAEs in only a quarter degree field (e.g., compared to
$\sim$10 QSOs at z$>$6 which are distributed over a quarter of the
sky! Fan et al.\ 2004). This implies that LAEs may play an important
role in reionizing the universe at z$>$6 (for a review see Fan,
Carilli \& Keating, 2006).  Investigating the physical properties of
these sources are thus of great interest and ALMA will play a critical
role in studying these objects, as discussed in the following.

\begin{figure}
\begin{center}
  \includegraphics[scale=0.45]{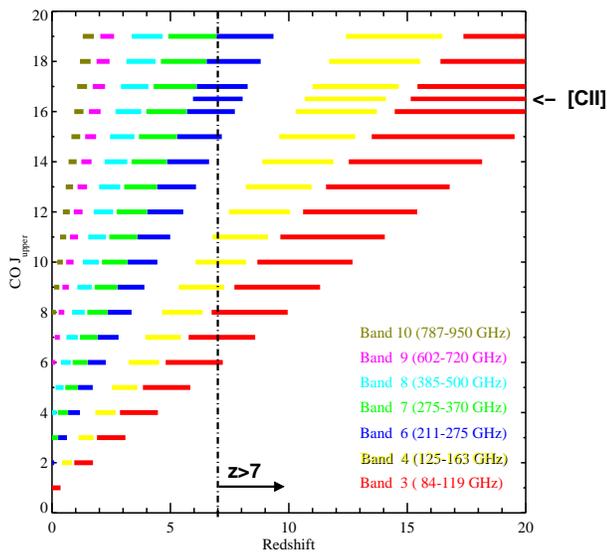}
\end{center}
\caption{ALMA CO `discocery space': The horizontal lines indicate
  which CO transition (plotted on the y--axis) can be observed with
  which ALMA band as a function of redshift (plotted on the x--axis).
  For objects with z$>$7, only the higher--J CO transitions can be
  observed with ALMA. The [CII] `discovery space' is also indicated.}
\label{fig:1}       % Give a unique label
\end{figure}

\section{Interstellar Medium: CO vs. [CII] emission}
\vspace*{-2.5mm}

\subsection{Carbon Monoxide (CO)}
\vspace*{-2.5mm}

Constraining the properties of the molecular gas in objects at the end
of cosmic reionization is clearly of key importance as such
observations 1) will measure the available 'fuel' for star formation,
2) will help to constrain the dynamical mass of the system and will
thus 3) allow to put these objects in an evolutionary context for
early galaxy formation.  Typically, at low and high redshifts, CO
emission is used as a tracer for the molecular gas phase (e.g. review
by Solomon \& Vanden Bout 2004). It is important to keep in mind
though, that, at the highest z, only the very high rotational lines of
CO will be observable with ALMA.  E.g. even in the lowest (currently
funded) frequency band of ALMA (band 3, 84--119\,GHz), only CO
transitions with J$>$7 (i.e., CO(7--6), CO(8--7), etc.) will be
observable at z$>$7. This is graphically illustrated in Fig.~1 where
we plot ALMA's `CO discovery space' (i.e., which line can be observed
at which redshift using which ALMA band). The high--J transitions
correspond to highly excited gas (either due to high kinetic
temperatures, high densities, or both) which may not be excited in
normal starforming environments.  This is shown in Fig.~2 (taken from
Weiss et al.\ 2005) where measured CO line strengths (as a function of
J, this is sometimes referred to as CO line ladders/SEDs) are plotted
for a number of key sources. What is immediately obvious from this
plot is that most objects have sharply decreasing CO line strengths
beyond J$>8$, in particular starforming systems such as NGC\,253, or
the sub--millimeter galaxy plotted in this diagram (the quasars appear
to be more excited, but their CO line SED still turns over at
J$\sim$7, for an exceptional object see APM\,07279, Weiss et al.\
2007). This comparison immediately implies that emission from the CO
molecule will typically be very difficult to observe with ALMA at
z$>$7 as the observable lines will simply not be excited.

\begin{figure}
\begin{center}
  \includegraphics[scale=0.45]{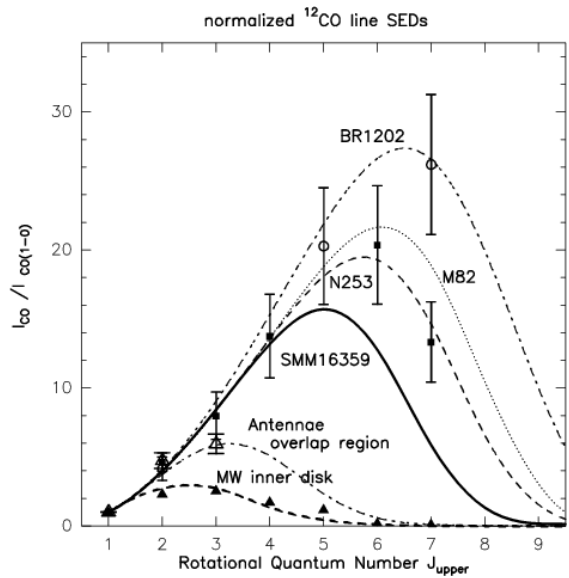}
\end{center}
\caption{Comparison of various normalized (by their CO(1--0) flux
  density) CO line SEDs at low and high redshift (figure taken from
  Weiss et al.\ 2005). The CO line SEDs decline rapidly beyond
  J=6--8.}
\label{fig:2}       % Give a unique label
\end{figure}

\subsection{Ionized Carbon ([CII]) to the Rescue!}
\vspace*{-2.5mm}

An alternative tracer of the interstellar medium is one of the main
cooling line of the ISM, the $^2$P$_{3/2}$ $\to$ $^2$P$_{1/2}$
fine--structure line of C$^+$ (or [CII]).  In brief, the [CII] line is
expected to be much stronger than any of the CO lines.  Given its high
frequency (157.74\,$\mu$m, corresponding to 1900.54\,GHz) [CII]
studies in the local universe are limited to airborne or satellite
missions (e.g.  Stacey et al.\ 1991, Malhotra et al.\ 1997, Madden et
al.\ 1997).  These studies have demonstrated that this single line can
indeed carry a good fraction of the total infrared luminosity (L$_{\rm
  FIR}$) of an entire galaxy.  In the local universe, the ratio
L$_{\rm CII}$/L$_{\rm FIR}$ has been found to be 2--5$\times$10$^{-4}$
in the case of ULIRGS (e.g. Gerin \& Phillips 2000), but is more like
5--10$\times$10$^{-3}$ in more typical starforming galaxies (for a
discussion on possible reasons for the supressed ratio in ULIRGs see,
e.g., Luhman et al.\ 1998). Notably, the ratio has been found to be
1\% or even higher in low metallicity environments. E.g., in the low-
metallicity galaxy IC10, L$_{\rm CII}$/L$_{\rm FIR}$ reaches values as
high as 4\%, with an average value of 2\% (Madden et al.\ 1997, see
Israel et al.\ 1996 for a similar result for the LMC). This is the
reason why it has long been argued (e.g., Stark 1997) that observation
of the [CII] line of prestine systems at the highest redshifts will
likely be the key to study molecular gas in the earliest starforming
systems, in particular in the era of ALMA. The ALMA [CII] `discovery
space' is also indicated in Fig.~1.

\begin{figure}
\begin{center}
  \includegraphics[scale=0.4]{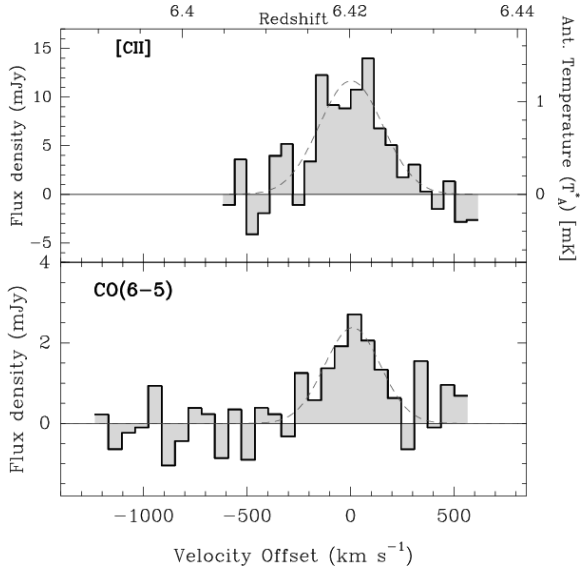}
\end{center}
\caption{{\em Top:} First detection of [CII] at high redshift in the
  z=6.42 QSO J1148+5251 (Maiolino et al.\ 2005). {\em Bottom:}
  brightest CO transition (J=6) in the same source (Bertoldi et al.\
  2003, Walter et al.\ 2003). Note that the [CII] line is brighter by a
  factor of $\sim$5.}
\label{fig:3}       % Give a unique label
\end{figure}

\section{Expected [CII] line strengths}
\vspace*{-2.5mm}

At the redshifts of the LAEs, the [CII] line is shifted to the 1\,mm
band of ALMA (band 6, 211--275\,GHz). [CII] emission has recently been
successfully detected using the IRAM 30\,m in the highest redshift
quasar J1148+5251 at z=6.42 (Maiolino et al.\ 2005, see Fig.~3).  The
noteable difference between J1148+5251 and the z$>$6 LAEs is that the
ratio L$_{\rm CII}$/L$_{\rm FIR}$ has been found to be very low
($\sim$5$\times$10$^{-4}$) in J1148+5251, i.e. in perfect agreement
with studies of low redshift ULIRGs that show a central AGN.  On the
contrary, the LAE are presumably pure starbursts (no evidence for an
AGN is found, Taniguchi et al.\ 2005) and they likely have lower
metallicities compared to the highly overdense regions in which the
luminous quasars are supposedly present. All these arguments point
towards a L$_{\rm CII}$/L$_{\rm FIR}$ ratio in LAE that is close to
what is found for nearby normal galaxies, or perhaps even for the
metal--poor dwarf galaxies (i.e., around 1\% or even higher). In other
words, the [CII] luminosity of the LAE may well be an order of
magnitude stronger (for a given IR luminosity) than what has been
found in the z=6.4 QSO. In the following we present a quick
back--of--the envelope calculation based on the detected [CII] line
strength in J\,1148+5251 ($\sim$10~mJy) which has a SFR of a few 1000
M$_\odot$\,yr$^{-1}$. This SFR is more than two order of magnitudes
higher than the SFR found in a typical LAE, but as the L$_{\rm
CII}$/L$_{\rm FIR}$ may be higher by an order of magnitude in the
LAEs, the expected [CII] line strength of the LAE may be as high as
1\,mJy. Such a line should be easily detectable with ALMA at high
significance in a few hours.

\section{Resolving the ISM}
\vspace*{-2.5mm}

{\em Detecting} the [CII] (or CO) emission is critical to estimate the
reservoir of the (molecular) gas in these early systems. A second step
is then to  spatially resolve the molecular gas distribution. In
particular, given the typical diameters of galaxies of many kpc, a
linear resolution of $\sim 1$\,kpc is needed to resolve the structure
of the underlying galaxy. Such measurements are needed 1) to get an
estimate for the size of the galaxy (and thus a better estimate
for the dynamical mass), 2) to resolve potentially merging systems and
3) to better constrain the physical properties of the gas (e.g.,
by measuring the brightness temperature of the hosts).  A linear
resolution of 1\,kpc corresponds to a resolution of 0.15$''$ at the
redshifts under consideration (1$''\sim5.8$\,kpc at $z=6$).  Such
observations can then in turn be used to constrain the predictions by
CDM simulations of early galaxy formation, and, if a large sample was
available, put limits on the frequency of mergers at high redshift. In
addition, such studies can be used to constrain the possible
redshift-evolution of the M$_{\rm BH}$--$\sigma_{\rm v}$ relation in
high--z quasars. Such observations will clearly be feasible with ALMA
in the extended arrays. High--resolution CO imaging is already
possible with the current generation of telescopes: we have used the
VLA to resolve the molecular gas in the host galaxy of the z=6.42 QSO
J\,1148+5251 (see Fig.~4, Walter et al.\ 2004).

\section{The case for ALMA band 5}
\vspace*{-2.5mm}

As a technical note: the ALMA redshift coverage for the [CII] line is
not ideal as its frequency lies between the CO(17--16) and CO(16--15)
transition (see Fig.~1). One concern is that the critical redshift
range (8$<$z$<$10.5) is currently not fully covered: This frequency
range corresponds to the ALMA band 5 which is only partly funded by
the European Union as part of the Sixth Framework Programme (FP6) for
up to 8 antennas. Clearly, it would be highly desireable to equip as
many ALMA antennas with band 5 receivers as possible.

\begin{figure}
\begin{center}
  \includegraphics[scale=0.5]{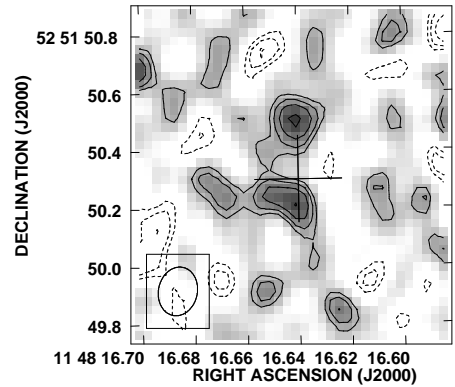}
\end{center}
\caption{High--resolution CO image of the z=6.42 QSO J\,1148+5251
  obtained at the VLA (Walter et al.\ 2004). The resolution achieved
  in these observations (0.15$''$, corresponding to $\sim$1\,kpc) will
  be routinely reached with ALMA.}
\label{fig:4}       % Give a unique label
\end{figure}

\section{Concluding remarks}

ALMA observations of the [CII] line will play a fundamental role in
studying the youngest galaxies in the Epoch of Cosmic Reionization at
z$>$7. Given the expected line strengths it should be possible to
resolve these galaxies in the [CII] line emission on kpc scales. Such
measurements would not only constrain the sizes but would also help to
derive the dynamical masses in these early starforming systems. Given
the typical CO excitation in starforming galaxies (i.e. drop in
excitation around the J$\sim$6 transitions), ALMA will likely act as a
[CII]-- rather than a CO--machine for objects at these extreme
redshifts.


\begin{thebibliography}{}


\bibitem{}Bertoldi, F., Carilli, C. L., Cox, P., Fan, X., et al.\
  2003, A\&A, 406, L55
\bibitem{}Fan, X., Hennawi, J.~F., Richards, G.~T., et al.\ 2004, AJ,
  128, 515
\bibitem{}Fan, X., Carilli, C.L., Keating, B., 2006, ARA\&A, 44, 415
\bibitem{}Gerin, M., \& Phillips, T.~G.\ 2000, ApJ, 537, 644
\bibitem{Hu02}Hu, E.~M., Cowie, L.~L., McMahon, R.~G., et al.\ 2002,
  ApJ, 568, L75
\bibitem{}Israel, F.~P., Maloney, P.~R., Geis, N., Herrmann, F.,
  Madden, S.~C., Poglitsch, A., Stacey, G.J., 1996, ApJ, 465, 738
\bibitem{}Iye, M., Ota, K., Kashikawa, N., Furusawa, H., Hashimoto,
  T., Hattori, T., Matsuda, Y., Morokuma, T., Ouchi, M., Shimasaku,
  K., 2006, Nature, 443, 186
\bibitem{Kurk04}Kurk, J.~D., Cimatti, A., di Serego Alighieri, et al.\
  2004, A\&A, 422, L13
\bibitem{}Luhman, M.~L., et al.\ 1998, ApJL, 504, L11
\bibitem{}Madden, S.~C., Poglitsch, A., Geis, N., Stacey, G.~J., \&
  Townes, C.~H.\ 1997, ApJ, 483, 200
\bibitem{}Maiolino, R., Cox, P., Caelli, P., et al.\ 2005,
  astro-ph/0508064
\bibitem{}Malhotra, S., et al.\ 1997, ApJL, 491, L27
\bibitem{}Murayama, T., Taniguchi, Y., Scoville, N.Z., et al.\ 2007, ApJS, astro-ph/0702458
\bibitem{}Stacey, G.~J., Geis, N., Genzel, R., Lugten, J.~B.,
  Poglitsch, A., Sternberg, A., \& Townes, C.~H.\ 1991, ApJ, 373, 423
\bibitem{}Stark, A.~A.\ 1997, ApJ, 481, 587
\bibitem{Stern05}Stern, D., Yost, S.~A., Eckart, M.~E., Harrison,
  F.~A., Helfand, D.~J., Djorgovski, S.~G., Malhotra, S., \& Rhoads,
  J.~E.\ 2005 , ApJ, 619, 12
\bibitem{}Taniguchi, Y., Ajiki, M., Nagao, T. et al.\ 2005, PASJ, 57,
  165
\bibitem{}Walter, F., Bertoldi, F., Carilli, C. L., et al.\ 2003,
  Nature, 424, 406
\bibitem{}Walter, F., Carilli, C., Bertoldi, F., et al.\ 2004, ApJ, 615, L17
\bibitem{}Weiss, A., Downes, D., Neri, R., Walter, F., Henkel, C.,
  Wilner, D.J., Wagg, J., Wiklind, T., 2007, A\&A, in press
  (astro--ph/0702669)


\end{thebibliography}


\end{document}